\documentclass[12pt]{article}

\usepackage{latexsym}
\usepackage{amsmath}
\usepackage{amssymb}
\catcode`\@=11
\newcount\hour
\newcount\minute
\newtoks\amorpm \hour=\time\divide\hour by 60\minute
=\time{\multiply\hour by 60 \global\advance\minute by-\hour}
\edef\standardtime{{\ifnum\hour<12 \global\amorpm={am}%
        \else\global\amorpm={pm}\advance\hour by-12 \fi
        \ifnum\hour=0 \hour=12 \fi
        \number\hour:\ifnum\minute<10
        0\fi\number\minute\the\amorpm}}
\edef\militarytime{\number\hour:\ifnum\minute<10
0\fi\number\minute}
\def\draftlabel#1{{\@bsphack\if@filesw {\let\thepage\relax
   \xdef\@gtempa{\write\@auxout{\string
      \newlabel{#1}{{\@currentlabel}{\thepage}}}}}\@gtempa
   \if@nobreak \ifvmode\nobreak\fi\fi\fi\@esphack}
        \gdef\@eqnlabel{#1}}
\def\@eqnlabel{}
\def\@vacuum{}
\def\marginnote#1{}
\def\draftmarginnote#1{\marginpar{\raggedright\scriptsize\tt#1}}
\def\draft{
        \pagestyle{plain}
        \overfullrule=2pt
        \oddsidemargin -.1truein
        \def\@oddhead{\sl \phantom{\today\quad\militarytime} \hfil
        \smash{\Large\sl DRAFT} \hfil \today\quad\militarytime}
        \let\@evenhead\@oddhead
        \let\label=\draftlabel
        \let\marginnote=\draftmarginnote
        \def\ps@empty{\let\@mkboth\@gobbletwo
        \def\@oddfoot{\hfil \smash{\Large\sl DRAFT} \hfil}
        \let\@evenfoot\@oddhead}
        \def\@eqnnum{(\theequation)\rlap{\kern\marginparsep\tt\@eqnlabel}%
        \global\let\@eqnlabel\@vacuum}  }

\renewcommand{\theequation}{\thesection.\arabic{equation}}
\renewcommand{\thefootnote}{\fnsymbol{footnote}}

\def\appendix#1{\addtocounter{section}{1}\setcounter{equation}{0}
\renewcommand{\thesection}{\Alph{section}}
\section*{Appendix \thesection\protect\indent \parbox[t]{11.15cm}{#1}}
\addcontentsline{toc}{section}{Appendix \thesection\ \ \ #1}}

\def\be{\begin{equation}}
\def\ee{\end{equation}}

\catcode`\@=11

\begin{document}
\date{October 2005}
\begin{titlepage}
\begin{center}
\vspace{4.0cm}
{\Large \bf The Classification of Highly Supersymmetric Supergravity Solutions}\\[.2cm]

\vspace{1.5cm}
 {\large  U.~Gran$^1$, J.~Gutowski$^2$, G.~Papadopoulos$^2$  and D.~Roest$^3$ }

\vspace{0.5cm}

${}^1$ Fundamental Physics\\
Chalmers University of Technology\\
SE-412 96 G\"oteborg, Sweden\\

\vspace{0.5cm}
${}^2$ Department of Mathematics\\
King's College London\\
Strand\\
London WC2R 2LS, UK\\

\vspace{0.5cm}
${}^3$ Centre for Theoretical Physics \\ University of Groningen \\
Nijenborgh 4, 9747 AG Groningen, The Netherlands

\end{center}
\vskip 1.5 cm
\begin{abstract}
The spinorial geometry method is an effective method for constructing
systematic classifications of supersymmetric supergravity solutions.
Recent work on analysing highly supersymmetric solutions in type IIB supergravity 
using this method is reviewed \cite{iibpreon, iibnearmax}. It is shown that
all supersymmetric solutions of IIB supergravity with more than 28 Killing spinors
are locally maximally supersymmetric.

\end{abstract}


\vskip2mm


\end{titlepage}
\newpage
\setcounter{page}{1}
\renewcommand{\thefootnote}{\arabic{footnote}}
\setcounter{footnote}{0}

\setcounter{section}{0}
\setcounter{subsection}{0}

\section{Introduction}

The classification of supersymmetric supergravity solutions is of
importance in the context of string theory; it is also of
intrinsic mathematical interest. The initial classification
was constructed for certain four-dimensional theories
in \cite{tod}. Following this work, the first classification
of supersymmetric solutions in a higher dimensional
theory was constructed in \cite{hull:2003}, where all
supersymmetric solutions of the minimal ungauged
$N=2$, $D=5$ supergravity theory were classified.
This was then extended to more 
complicated five-dimensional theories \cite{gauntlett:2004,
gauntlett2:2005, sabra:2005}. These
classifications were used to investigate black
holes in five dimensions; in particular,
they were used to find a
supersymmetric black ring \cite{reall:2004},
as well as the first example of a ``black Saturn"
solution \cite{gauntlett1:2005}, \cite{gauntlett2:2005},
consisting of an arbitrary number of concentric
black rings, with a black hole at the centre.
The first regular, asymptotically $AdS_5$, black holes
were also found in \cite{adsbh} using the
five-dimensional classifications; these solutions
have since been further generalized \cite{pope:2005, kunduri:2006}.

These low-dimensional classifications were constructed by
investigating the properties of differential forms 
obtained from bi-linears in the Killing spinors.
These forms satisfy algebraic constraints, which
follow from the Fierz identities, and also differential
constraints arising from the Killing spinor equations. 
The constraints imply the existence of symmetries and
different types of geometric structures
together with constraints on the various fluxes which
appear in the theories.
This method was successfully used to find solutions
preserving low proportions of supersymmetry
in low-dimensional supergravities, and also
in $D=11$ supergravity \cite{pakis1:2003, pakis2:2003}.
However, is not particularly well adapted for investigating
more complicated theories, such as type IIB supergravity.
Furthermore, it is difficult to use it to classify, in a systematic
fashion, solutions preserving higher proportions of supersymmetry.
It is, however, possible to construct restricted classifications of solutions
if one assumes that the solutions have 
additional constraints on the spacetime geometry,
such as product structures involving $AdS$ geometries \cite{gauntlettads1,
gauntlettads2, gauntlettads3}, a number of which are of interest in the context of the
AdS/CFT correspondence. 

Recently, considerable further progress has been made using techniques
of spinorial geometry.
This was originally formulated to classify solutions
of $D=11$ and type IIB supergravity \cite{papadgran1,
papadgran2, papadgran3}, and has also been applied to heterotic and type I supergravity
\cite{papadheterotic, papadtype1},
as well as various supergravity theories in lower dimensions 
\cite{n2d4, n1d4, n1ortin, fivespin1, fivespin2}.
In applying the spinorial geometry method to
a particular theory, the first step is to
write the spinors as differential forms \cite{spinform1, spinform2} 
and determine the action of the Cifford algebra generators on the spinors,
working in an appropriately chosen basis.
Then, one applies gauge transformations to the
spinors, in order to reduce them to simplified canonical forms.
The simplified spinors are then substituted into the
Killing spinor equations, or their integrability conditions,
and constraints on the spacetime geometry and fluxes
are thereby obtained.

The spinorial geometry method 
allows one, for the first time,  to perform a complete and systematic classification
of supersymmetric solutions preserving large amounts of
supersymmetry, without imposing any additional assumptions
on the spacetime structure.
In this context, it is notable that 
a complete classification of maximally supersymmetric solutions
of type II and $D=11$ supergravity has been constructed only
comparatively recently \cite{papamaximal1}.
A natural progression of this analysis is to consider solutions
of type II and $D=11$ supergravity 
preserving the next to maximal proportion of the supersymmetry,
i.e. for which the space of Killing spinors is 31-dimensional.
Such hypothetical solutions, termed {\it preons},
were considered in \cite{bandospreon1, bandospreon2},
although no explicit solutions were found. Partial
non-existence theorems were constructed
\cite{duffnogo}, but the status of preons remained
unresolved for some time \cite{hullnogo}.
Using spinorial geometry techniques, it has been
proven that there are no preons in IIB supergravity \cite{iibpreon}, and
also no preons  in $D=11$ supergravity \cite{d11preon, figquot}.
In this review we will concentrate on highly supersymmetric
solutions of IIB supergravity. We review the non-existence
theorem for preons formulated in \cite{iibpreon},
and also an extension of this result found in 
\cite{iibnearmax}, in which is is shown that there
are no solutions of IIB supergravity preserving 30/32
and 29/32 of the supersymmetry either.

The plan of this review is as follows. In Section 2 we review 
the Killing spinor equation of IIB supergravity, and the construction
of canonical forms for spinors as differential forms. In Section
3, we present a proof that all preon solutions of IIB supergravity
are locally maximally supersymmetric \cite{iibpreon}.
In Section 4 we prove that all
solutions preserving 30/32 of the supersymmetry are
locally maximally supersymmetric,
and in Section 5, we prove that all solutions preserving
29/32 of the supersymmetry are locally maximally supersymmetric
\cite{iibnearmax}.
In Section 6 we present our conclusions.

\section{Killing Spinors in IIB Supergravity} \label{}

In this section, we review the formulation of the Killing spinor
equation given in \cite{schwarziib}. We also present the
construction of canonical forms for Killing spinors given in 
\cite{papadgran2}.

\subsection{The Killing Spinor Equations}

The bosonic field content of IIB supergravity consists of the spacetime metric $g$, two real scalars called the axion
$\sigma$ and dilaton $\phi$ , two complex 3-form field strengths $G^\alpha=d A^\alpha$ ($\alpha=1,2$)
satisfying $G^1 = (G^2)^*$, and
a real self-dual 5-form field strength $F$

\begin{eqnarray}
F_{M_1M_2M_3M_4M_5}=5\partial_{[M_1} A_{M_2M_3M_4M_5]}+{5i\over8} \epsilon_{\alpha\beta} A^\alpha_{[M_1M_2} G^\beta_{M_3M_4M_5]}~,
\end{eqnarray}
where $\epsilon^{12}=1=\epsilon_{12}$.
The gradients of the axion and dilaton are combined into a 1-form $P$,
and the 3-forms $G^1, G^2$ are combined into a complex 3-form $G$. This
is achieved by introducing an $SU(1,1)$ matrix 
\begin{equation}
\left(
\begin {array}{cc}
V^1_-&V^1_+\\
\noalign{\medskip}
V^2_-&V^2_+
\end {array}
\right)
\end{equation}
where the components $V^\alpha_\pm$ are constrained by
\begin{equation}
V_-^\alpha V_+^\beta-V_-^\beta V_+^\alpha=\epsilon^{\alpha\beta}~, \quad (V_-^1)^*=V_+^2, \qquad (V_-^2)^*=V_+^1 \ .
\end{equation}
$V^\alpha_\pm$ are then fixed in terms of the axion and dilaton
by
\begin{equation}
{V_-^2 \over V_-^1}={1+i(\sigma+i e^{-\phi}) \over 1-i(\sigma+i e^{-\phi})}~.
\end{equation}
Then the complex 1-form $P$ and complex 3-form $G$ are defined by
\begin{equation}
P_M=-\epsilon_{\alpha\beta} V_+^\alpha \partial_M  V_+^\beta , \quad G_{MNR}=-\epsilon_{\alpha\beta} V^\alpha_+  G_{MNR}^\beta \ .
\end{equation}

A bosonic solution of IIB supergravity preserves a proportion of  supersymmetry if it admits
a Killing spinor $\epsilon$ satisfying the gravitino and the algebraic Killing spinor equations.

The gravitino Killing spinor equation is: 
\begin{eqnarray}
\label{graveq}
\tilde\nabla_M \epsilon+{i\over 48} \Gamma^{N_1\dots N_4 } \epsilon
 F_{N_1\dots N_4 M} 
 -{1\over 96} (\Gamma_{M}{}^{N_1N_2N_3}
G_{N_1N_2N_3}  -9 \Gamma^{N_1N_2} G_{MN_1N_2}) (C\epsilon)^*=0
\end{eqnarray}
where
\begin{equation}
\tilde\nabla_M=\partial_M-{i\over2}Q_M +{1\over4} \Omega_{M,AB} \Gamma^{AB}
\end{equation}
is the standard covariant derivative
twisted with $U(1)$ connection $Q_M$, given in terms of the $SU(1,1)$ scalars by
\begin{equation}
Q_M=-i \epsilon_{\alpha\beta} V_-^\alpha\partial_M V_+^\beta
\end{equation}
and $\Omega$ is the spin connection.

The algebraic Killing spinor equation is a purely algebraic constraint:
\begin{equation}
\label{dileq}
P_M \Gamma^M (C\epsilon)^*+ {1\over 24} G_{N_1N_2N_3} \Gamma^{N_1N_2N_3} \epsilon=0 \ .
\end{equation}
The Killing spinor $\epsilon$ which appears in these equations is a complex Weyl spinor
\begin{equation}
\epsilon= \eta_1 +i \eta_2
\end{equation}
where $\eta_1, \eta_2$ are Majorana-Weyl spinors. The space of Majorana-Weyl
spinors is denoted by  $\Delta_{16}^+$; Majorana-Weyl spinors $\eta$ satisfy
\begin{equation}
\eta = C (\eta^*)
\end{equation}
and $C$ is the charge conjugation matrix with the property that
\begin{equation}
C^{-1} \Gamma_M C = (\Gamma_M)^* \ .
\end{equation}
A basis can be chosen in which $C= \Gamma_{6789}$.

\subsection{Spinors as Differential Forms}

In order to analyse the solutions of the Killing spinor
equations, we shall take the spinors to be differential forms.
To formulate this construction, first take
$e_1, \dots, e_5$ to be a locally defined orthonormal basis of ${\mathbb{R}}^{5}$,
and let $U$ be the span of $e_1, \dots, e_5$ over ${\mathbb{R}}$. Then the space
of Dirac spinors $\Delta_c$ is the complexified space of all differential
forms over $U$;
\begin{equation}
\Delta_c = \Lambda^*(U \otimes {\mathbb{C}}) \ .
\end{equation}
$\Delta_c$ decomposes into even forms $\Delta_c^+$ and odd forms $\Delta_c^-$,
which are the complex Weyl representations of $Spin(9,1)$.
The gamma matrices are represented on $\Delta_c$ as
\begin{eqnarray}
\Gamma_0\eta&=& -e_5\wedge\eta +e_5 \lrcorner\eta
\cr
\Gamma_5\eta&=& e_5\wedge\eta+e_5 \lrcorner \eta
\cr
\Gamma_j\eta&=& e_j\wedge \eta+ e_j \lrcorner \eta \qquad \qquad j=1,\dots,4
\cr
\Gamma_{5+j}\eta&=& i e_j\wedge\eta-ie_j \lrcorner\eta \qquad \quad \ \ j=1,\dots,4 \ .
\end{eqnarray}

In order to investigate highly supersymmetric solutions, it is necessary to introduce
a gauge-invariant inner product. First, define an inner product $<,>$ on complexified 1-forms via
\begin{equation}
<z^a e_a, w^b e_b>=\sum_{a=1}^{5}  (z^a)^* w^a 
\end{equation}
for $z^a, w^b \in {\mathbb{C}}$. This inner product is then extended  onto the whole of $\Delta_c$.
The gamma matrices are defined in such a way that $\Gamma_j$ for $j=1, \dots, 9$ are hermitian and
$\Gamma_0$ is anti-hermitian with respect to this inner product. However, $<,>$ is not 
$Spin(9,1)$ gauge invariant. It is, however, straightforward to define
a $Spin(9,1)$ invariant inner product $B$ on $\Delta_c$, given by
\begin{equation}
B(\epsilon_1,\epsilon_2)= < \Gamma_0 C (\epsilon_1)^*, \epsilon_2> \ .
\end{equation}
$B$ is skew-symmetric in $\epsilon_1, \epsilon_2$, and vanishes when
restricted to $\Delta^+_c$ or $\Delta^-_c$.
Using $B$, one can then define a non-degenerate pairing ${\cal{B}} : \Delta^+_c \otimes \Delta^-_c \rightarrow {\mathbb{R}}$
given by
\begin{equation}
{\cal{B}} (\epsilon, \xi) = {\rm Re \ } B(\epsilon,\xi) \ .
\end{equation}

\subsection{Canonical Forms for Spinors}

A key step in the application of spinorial geometry techniques to the analysis 
of supersymmetric solutions is the simplification of spinors using gauge transformations.

First, consider a Majorana-Weyl spinor $\eta_1 \in \Delta_{16}^+$. 
It has been shown that $Spin(9,1)$ has one type of orbit with stability subgroup $Spin(7)\ltimes {\mathbb{R}}^8$ in $\Delta_{16}^+$
\cite{bryant, hernandez, mwave}. To prove this, one decomposes $\Delta_{16}^+$ as 
\begin{equation}
\Delta_{16}^+= {\mathbb{R}}<1+e_{1234}>+\Lambda^1({\mathbb{R}}^7)+\Delta_8~,
\end{equation}
where
${\mathbb{R}}<1+e_{1234}>$ is the singlet generated by $1+e_{1234}$,
$\Lambda^1({\mathbb{R}}^7)$ is the vector representation of $Spin(7)$ spanned by Majorana spinors
associated with  2-forms in the directions $e_1, e_2, e_3, e_4$ and by $i(1-e_{1234})$; 
$\Delta_8$ is the spin representation of $Spin(7)$ spanned by the remaining Majorana spinors
of type $e_5 \wedge \eta'$ where $\eta'$ is generated by odd forms in the directions
$e_1, e_2, e_3, e_4$.

In order to simplify $\eta_1$, recall that  $Spin(7)$ acts transitively on the $S^7$ in $\Delta_8$,
with stability subgroup $G_2$, and
$G_2$ acts transitively on the $S^6$ in $\Lambda^1 ({\mathbb{R}}^7)$ with stability subgroup $SU(3)$
\cite{salamon}.
Using these actions, one can show that $\eta_1$ 
lies in the orbit of $1+e_{1234}$. This spinor is $Spin(7)\ltimes {\mathbb{R}}^8$ invariant.
To see this, it is convenient to work in a hermitian basis, with gamma matrices
\begin{equation}
\Gamma_{\bar\alpha}= {1\over \sqrt {2}}(\Gamma_\alpha+i \Gamma_{\alpha+5})~,~~~
\Gamma_\pm={1\over \sqrt{2}} (\Gamma_5\pm\Gamma_0)
~,~~~\Gamma_{\alpha}= {1\over \sqrt {2}}(\Gamma_\alpha-i \Gamma_{\alpha+5})~
\end{equation}
for $\alpha, \beta=1,2,3,4$, and with metric
\begin{equation}
ds^2 = 2 e^+ e^- + 2 \delta_{\alpha \bar{\beta}} e^\alpha e^{\bar{\beta}} \ .
\end{equation}
The spinor $\eta_1$ can then be written as
\begin{equation}
\eta_1= a(1+e_{1234})+ \theta_1+\theta_2~,
\end{equation}
with $a \in {\mathbb{R}}$, $\theta_1\in \Lambda^1({\mathbb{R}}^7)$ and $\theta_2\in \Delta_8$ There are several cases to consider:

\begin{itemize}

\item[i)] $a \neq 0$, $\theta_2=0$. Using the transitive action of $G_2 \subset Spin(7)$ on the $S^6$ in $\Lambda^1 ({\mathbb{R}}^7)$,
make a gauge transformation so that $\theta_1=i b(1-e_{1234})$, and hence
\begin{equation}
\eta_1= a (1+e_{1234})+ i b (1-e_{1234}) = \sqrt{a^2+b^2} e^{\arctan ({b \over a}) \Gamma_{16}} (1+e_{1234}) \ .
\end{equation}
So $\eta_1$ lies in the same orbit as $1+e_{1234}$.

\item[ii)] $a \neq 0$, $\theta_2 \neq 0$. Using the transitivity of the action of $Spin(7)$ on the $S^7$ in ${\mathbb{R}}^8$, make a gauge transformation and set
$\theta_2 = c \Gamma^+ (e_1 + e_{234})$
Also, using the transitivity of the action of $G_2$ on the $S^6$ in $\Lambda^1 ({\mathbb{R}}^7)$,
a gauge transformation can be chosen so that $\theta_1=i b (1-e_{1234})$, and $\theta_2$ is unaffected.
Then
\begin{eqnarray}
\eta&=&a (1+e_{1234})+ i b (1-e_{1234})+ c \Gamma^+ (e_1+ e_{234}) 
\cr
&=& e^{{b\over 2c} \Gamma^- \Gamma^6} e^{{c\over a} \Gamma^+ \Gamma^1}
a (1+e_{1234}) \ .
\end{eqnarray}
Again, $\eta_1$ lies in the same orbit as $1+e_{1234}$.

\item[iii)] $a=0$. This orbit is represented by $c \Gamma^+ (e_1+e_{234})$, which is also in
the  orbit of $1+e_{1234}$, using the action of the $\Gamma_{51}$ generator of $Spin(9,1)$.

\end{itemize}

So in all possible cases, by making a gauge transformation, one can set
\begin{equation}
\eta_1 = f (1+e_{1234}) \ .
\end{equation}
Having simplified the structure of  $\eta_1$, it is then straightforward to
write $\epsilon=\eta_1+i \eta_2$ in a simplified canonical form;
one applies $Spin(7) \ltimes {\mathbb{R}}^8 $ gauge transformations
to $\epsilon$, which leave $\eta_1$ invariant, and are chosen
to simplify $\eta_2$ as much as possible.

By using $Spin(7)$ gauge transformations, which leave
$\eta_1$ invariant, one can write
\begin{equation}
\eta_2= b_1 (1+e_{1234})+ i b_2 ( 1-e_{1234})+ b_3 (e_{15}+e_{2345}) \ .
\end{equation}
There are again various cases to consider.
\begin{itemize}

\item[i)] If $b_3 \neq 0$, then note that there exists a ${\mathbb{R}}^8$ transformation such that
\begin{equation}
\eta_2=e^{-{b_1\over2 b_3} \Gamma^-\Gamma^6+{b_2\over2 b_3} \Gamma^-\Gamma^1}
b_3 \Gamma^+ (e_1+e_{234})~.
\end{equation}
Hence, by using a $Spin(7) \ltimes {\mathbb{R}}^8$ gauge transformation, one can take
\begin{equation}
\eta_2= g (e_{15}+e_{2345}) \ .
\end{equation}
The stability subgroup of $Spin(9,1)$ which leaves $\eta_1$ and $\eta_2$ invariant is $G_2$.

\item[ii)] If $b_3=0$ then
\begin{equation}
\eta_2= g_1 (1+e_{1234})+ i g_2 ( 1-e_{1234})
\end{equation}
and the stability subgroup  of $Spin(9,1)$ which leaves $\eta_1$ and $\eta_2$ invariant is $SU(4)\ltimes {\mathbb{R}}^8$.

\item[iii)] If $b_2=b_3=0$ then
\begin{equation}
\eta_2 = g(1+e_{1234})
\end{equation}

and the stability subgroup of $Spin(9,1)$ which leaves $\eta_1$ and $\eta_2$ invariant is $Spin(7)\ltimes {\mathbb{R}}^8$.

\end{itemize}

To summarize, there are three canonical forms for a single even complex Weyl spinor $\epsilon \in \Delta^+_c$:
 \begin{eqnarray}
 Spin(7) \ltimes {\mathbb{R}}^8: && \epsilon =(f+ig)(1+e_{1234}) \,, \cr
 SU(4) \ltimes {\mathbb{R}}^8: && \epsilon=(f+ig_1-g_2)1 + (f+ig_1+g_2)e_{1234} \,, \cr
 G_2: && \epsilon = f(1+e_{1234}) + ig(e_{15}+e_{2345})\, . \label{normals}
 \end{eqnarray}
Furthermore, a directly analogous computation can be used to reduce
a single odd complex Weyl spinor $\nu \in  \Delta^-_c$ to
one of three canonical forms, with stability subgroups $Spin(7) \ltimes {\mathbb{R}}^8$,
$SU(4) \ltimes {\mathbb{R}}^8$ or $G_2$:
 \begin{eqnarray}
 Spin(7) \ltimes {\mathbb{R}}^8: && \nu =(n+im) (e_5+e_{12345}) \,, \cr
 SU(4) \ltimes {\mathbb{R}}^8: && \nu =(n-\ell+im) e_5+ (n+\ell+im) e_{12345} \,, \cr
 G_2: && \nu = n(e_5+e_{12345})+i m (e_1+e_{234}) \, . \label{oddnormals}
 \end{eqnarray}

\section{Solutions with $N=31$ Killing Spinors}

In this section, we shall prove that there are no solutions of type IIB supergravity which admit  exactly 31 
linearly independent Killing spinors. We recall that the maximally supersymmetric
solutions, i.e. those which have 32 linearly independent Killing spinors, have been
fully classified in \cite{papamaximal1}. There are only three maximally supersymmetric solutions:
${\mathbb{R}}^{9,1}$ (with $F=0, P=0, G=0$), $AdS_5 \times S^5$ (with $P=0, G=0$ but $F \neq 0$),
and a maximally supersymmetric plane wave solution (which has $P=0, G=0$ but $F \neq 0$).

In order to construct a non-existence theorem for preons in IIB supergravity;
suppose that there exists a solution with exactly (but no more than) 31
linearly independent Killing spinors over ${\mathbb{R}}$. Denote these Killing spinors
by $\epsilon^r$, for $r=1, \dots, 31$. The space of Killing spinors spanned by
the $\epsilon^r$ is orthogonal to a single {\it normal spinor}, $\nu \in \Delta_c^-$ with respect to the $Spin(9,1)$ invariant inner product ${\cal B}$. Using the results of the previous section, this normal spinor can
be brought into one of three simple canonical forms using $Spin(9,1)$ gauge transformations,
given in ({\ref{oddnormals}}).

One can write
 \begin{equation}
 \epsilon^r = \sum_{i=1}^{32} f^r{}_i \eta^i
 \end{equation}
 where $f^r{}_i$ are real, $\eta^p$ for $p=1, \dots ,  16$ is a basis for $\Delta^+_{16}$
 and $\eta^{16+p}=i\eta^p$.
The matrix with components $f^r{}_i$ is of rank 31, as
the functions $f^r{}_i$ are constrained by the orthogonality condition,
\begin{equation}
\label{orthog1}
{\cal{B}} (\epsilon^r , \nu )=0
\end{equation}
for $r=1, \dots , 31$.

There are three cases to consider. In the first case, $\nu =(n+im) (e_5+e_{12345})$
is $Spin(7) \ltimes {\mathbb{R}}^8$ invariant. One writes the Killing spinors as
\begin{eqnarray}
\epsilon^r= f^r{}_1 (1+e_{1234})+ f^r{}_{17} i (1+e_{1234})+ f^r{}_k
\eta^k \,,
\end{eqnarray}
where here $\eta^k$ denote the remaining basis elements of $\Delta^+_c$,
complementary to $1+e_{1234}$ and $i(1+e_{1234})$. On substituting the
spinors $\epsilon^r$ into ({\ref{orthog1}}), one obtains the constraint
\begin{eqnarray}
f^r{}_1 n- f^r{}_{17} m=0~.
\end{eqnarray}
Without loss of generality, one can take $n \neq 0$, 
and eliminate $f^r{}_1$ to obtain
\begin{eqnarray}
\epsilon^r={f^r{}_{17}\over n} (m+in) (1+e_{1234})+ f^r{}_k \eta^k~.
\end{eqnarray}
Similarly, for the cases when $\nu$ is  $SU(4) \ltimes {\mathbb{R}}^8$ and $G_2$ invariant,
one finds
\begin{eqnarray}
&&\epsilon^r={f^{r}{}_{17}\over n}[
(m+in)(1+e_{1234})]+{f^{r}{}_{18}\over n} [\ell (1+e_{1234})-n
(1-e_{1234})]+ f^{r}{}_k \eta^{k} \,,
\cr
&&\epsilon^r={f^{r}{}_{19}\over n} [m (1+e_{1234})+i n(e_{15}+e_{2345})]+ f^r{}_k \eta^k~,
\end{eqnarray}
where in each case $\eta^k$ denote the remaining basis elements for $\Delta^+_c$
which are orthogonal to $\nu$ and which do not depend on the functions various $m, n, \ell$
which appear in $\nu$. On substituting these spinors into
the algebraic constraint ({\ref{dileq}}), and using the fact that the matrix $f^r{}_i$ is of rank 31,
one obtains the following constraints when $\nu$ is $Spin(7) \ltimes {\mathbb{R}}^8$-invariant:
\begin{eqnarray}
&&P_M\Gamma^M C*[(m+in) (1+e_{1234})]+{1\over24} G_{M_1M_2M_3}\Gamma^{M_1M_2M_3} (m+in) (1+e_{1234})=0~,
\cr
&&P_M\Gamma^M\eta^p=0~,~~~G_{M_1M_2M_3}\Gamma^{M_1M_2M_3}\eta^p=0~,~~~p=2,\dots, 16~ .
\label{cona}
\end{eqnarray}
In the $SU(4) \ltimes {\mathbb{R}}^8$ case, one finds
\begin{eqnarray}
&&P_M\Gamma^M C*[(m+in) (1+e_{1234})]+{1\over24} G_{M_1M_2M_3}\Gamma^{M_1M_2M_3} (m+in) (1+e_{1234})=0~,
\cr
&&P_M\Gamma^M C*[\ell  (1+e_{1234})-n (1-e_{1234})]
\cr
&&~~~~~~~~~~+{1\over24} G_{M_1M_2M_3}\Gamma^{M_1M_2M_3} [\ell (1+e_{1234})-n (1-e_{1234})]=0~,
\cr
&&P_M\Gamma^M C*[i (1-e_{1234})]+{1\over24} G_{M_1M_2M_3}\Gamma^{M_1M_2M_3} [i (1-e_{1234})]=0~,
\cr
&&P_M\Gamma^M\eta^{p}=0~,~~~G_{M_1M_2M_3}\Gamma^{M_1M_2M_3}\eta^{p}=0~,~~~p=3,\dots,16~,
\label{conb}
\end{eqnarray}
and in the $G_2$ invariant case one finds
\begin{eqnarray}
&&P_M\Gamma^M C*[m (1+e_{1234})+in(e_{15}+e_{2345})]
\cr
&&~~~~~~~~~~~~~~+{1\over24} G_{M_1M_2M_3}\Gamma^{M_1M_2M_3} [m (1+e_{1234})+in(e_{15}+e_{2345})]=0~,
\cr
&&P_M\Gamma^M C*(i (1+e_{1234})+{1\over24} G_{M_1M_2M_3}\Gamma^{M_1M_2M_3} (i (1+e_{1234})=0~,
\cr
&&P_M\Gamma^M C* (e_{15}+e_{2345})+{1\over24} G_{M_1M_2M_3}\Gamma^{M_1M_2M_3} (e_{15}+e_{2345})=0~,
\cr
&&P_M\Gamma^M\eta^p=0~,~~~G_{M_1M_2M_3}\Gamma^{M_1M_2M_3}\eta^p=0~,~~~p=2,4,\dots,16~,
\label{conc}
\end{eqnarray}
where again, in each case, $\eta^p$ denote even Majorana-Weyl spinors,
which do not depend on the functions $m, n, \ell$, and are
such that the pair $(\eta^p, i \eta^p)$ are basis elements of $\Delta^+_c$ which are orthogonal to $\nu$.
On substituting these particular basis elements into ({\ref{dileq}}), and noting that
the presence of the operator $C*$ in this equation induces a relative minus sign
when evaluated on $\eta^p$ and $i \eta^p$, one finds that ({\ref{dileq}}) factorizes when
evaluated on these basis elements.

Consider the constraint
\begin{equation}
\label{pcon1}
P_M\Gamma^M\eta^p=0
\end{equation}
evaluated on these spinors. In particular, in all cases, one can take
\begin{equation}
\eta^p \in \{ e_{\alpha_1 \alpha_2} -{1 \over 2} \epsilon_{\alpha_1 \alpha_2 \beta_1 \beta_2} e_{\beta_1 \beta_2},
i( e_{\alpha_1 \alpha_2} +{1 \over 2} \epsilon_{\alpha_1 \alpha_2 \beta_1 \beta_2} e_{\beta_1 \beta_2}):
\alpha_1, \alpha_2, \beta_2, \beta_2 =1,2,3,4 \}
\end{equation}
and on evaluating ({\ref{pcon1}}) acting on these spinors, one finds that all components of $P$ with
the exception of $P_-$ are constrained to vanish. Next, taking
\begin{equation}
\eta^p \in \{  e_{\alpha 5} + {1 \over 6} \epsilon_{\alpha \beta_1 \beta_2 \beta_3} e_{\beta_1 \beta_2 \beta_3 5},
i ( e_{\alpha 5} - {1 \over 6} \epsilon_{\alpha \beta_1 \beta_2 \beta_3} e_{\beta_1 \beta_2 \beta_3 5}): \alpha=2,3,4 \}
\end{equation}
for all three cases, and evaluating ({\ref{pcon1}}) on these spinors, one finds that $P_-=0$ also.
Hence, for all three possible normal spinors $\nu$, one obtains the constraint
\begin{equation}
\label{shorter1}
P=0 \ .
\end{equation}
It follows directly that ({\ref{dileq}}) implies that $G=0$; this is because the constraint
$P=0$ implies that ({\ref{dileq}}) is linear over ${\mathbb{C}}$. Hence, if there are 31 linearly independent
solutions to ({\ref{dileq}}), there must be 32 linearly independent solutions, i.e.
\begin{equation}
G_{M_1M_2M_3}\Gamma^{M_1M_2M_3}\eta =0
\end{equation}
for all $\eta \in \Delta^c_+$. This forces all components of $G$ to vanish.

Hence, we have shown that type IIB supergravity preons must have $P=0$ and $G=0$.
To complete the analysis, consider the gravitino Killing spinor equation ({\ref{graveq}}).
As $P=0$, $G=0$, this equation is linear over ${\mathbb{C}}$. Therefore, if it has 31 linearly independent
solutions, it must in fact admit 32 linearly independent solutions. It follows that the
solution must be locally isometric to one of the maximally supersymmetric solutions.

\section{Solutions with $N=30$ Killing Spinors}

In this section, we review the analysis in \cite{iibnearmax}, in
which it is shown all solutions of type IIB supergravity
preserving 30/32 of the supersymmetry are locally isometric to
maximally supersymmetric solutions.
To prove this, we make use of a result
found in \cite{homogeneous1}, in which it is shown that all
solutions of type IIB supergravity preserving more than 24/32 of the
supersymmetry are homogeneous, and moreover, the ten linearly
independent Killing vectors are symmetries of the full solution.
In particular, the Lie derivative of the axion and the dilaton
with respect to these Killing vectors vanishes. Hence, for these solutions,
one finds that 
\begin{equation}
P=0 \ .
\end{equation}
We remark that the proof given in  \cite{homogeneous1} was
constructed after the non-existence proof for preons was
constructed in \cite{iibpreon}. It is clear that for preonic
solutions preserving 31/32 of the supersymmetry,
the constraint $P=0$ obtained from \cite{homogeneous1}
immediately implies that such solutions are excluded,
as a consequence of the reasoning following ({\ref{shorter1}})
at the end of the previous section.

	We begin by considering the case of 
solutions preserving 30/32 of the supersymmetry. 
Assuming that such solutions exist, we must have $P=0$,
and so the algebraic constraint ({\ref{dileq}}) simplifies to
\begin{equation}
 {1\over 24} G_{N_1N_2N_3} \Gamma^{N_1N_2N_3} \epsilon=0 \ .
 \end{equation}
 Note that this constraint is linear over ${\mathbb{C}}$, and so the Killing spinors 
 which satisfy this equation must be orthogonal to a normal spinor $\nu \in \Delta^c_-$
 with respect to the inner product $B$.  Again, $\nu$ can be brought into 
 one of three simple canonical forms ({\ref{oddnormals}}) using $Spin(9,1)$
 gauge transformations.
The solutions to the algebraic Killing spinor equation are
\begin{eqnarray}
\epsilon^r=\sum_{s=1}^{15} z^r{}_s\eta^s~,
\end{eqnarray}

where $\eta^i$ is an appropriately chosen basis normal to $\nu$ and $z$ is an invertible $15\times 15$ matrix of spacetime dependent complex functions. The algebraic Killing spinor constraint ({\ref{dileq}}) is then equivalent to
\begin{equation}
\label{akse}
 {1\over 24} G_{N_1N_2N_3} \Gamma^{N_1N_2N_3} \eta^s=0 \ .
 \end{equation}
There are three cases to consider, corresponding to the types of normal spinor $\nu$.
In all cases, one can choose the basis $(\eta^i)$ to have 13 (very simple) common elements, which
are orthogonal to $\nu$: $e_{pq}, e_{15pq}, e_{1p}, e_{1q}$ for $p=2,3,4$ and  $e_{15}-e_{2345}$.
Substituting these basis elements into the algebraic Killing spinor equation (\ref{akse}), we find that
the non-vanishing components of $G$ satisfy
 \begin{align}
  & G_{m \bar 1 \bar m} = - \tfrac12 G_{\bar 2 \bar 3 \bar 4} \,, \quad
  G_{- + \bar 1} = \tfrac12 G_{\bar 2 \bar 3 \bar 4} \,, \quad
  G_{+ 1 \bar 1} = G_{+ m \bar m} \,, \notag \\
  & G_{1 m \bar m} = - \tfrac12 G_{234} \,, \quad
  G_{- + 1} = \tfrac12 G_{234} \,,
 \end{align}
where $m = 2,3,4$, and there is no summation in the repeated $m$ indices. All other components of $G$ 
vanish.
The remaining two basis elements are case-dependent on the type of normal spinor $\nu$:
\begin{eqnarray}
 Spin(7) \ltimes {\mathbb{R}}^8: &&1-e_{1234}, e_{15}+e_{2345} \,, \cr
 SU(4) \ltimes {\mathbb{R}}^8: && e_{15}+e_{2345}, (n-\ell+im)1-(n+\ell+im)e_{1234}\,, \cr
 G_2: && 1-e_{1234}, m(1+e_{1234})+in(e_{15}+e_{2345}) \ .
 \end{eqnarray}
In all three cases, substituting the remaining basis elements into ({\ref{akse}}), one obtains the conditions
\begin{equation}
G_{234} = 0, \qquad  G_{\bar 2 \bar 3 \bar 4}, \qquad G_{+ 1 \bar 1}=0
\end{equation}
 which is sufficient to constrain all components of $G$ to vanish. 
 
 It remains to consider the integrability conditions of the Killing spinor equations 
for solutions with $G=P=0$. For such backgrounds,
the curvature ${\cal R}=[{\cal D}, {\cal D}]$ of the covariant connection ${\cal D}$
 of IIB supergravity can be expanded  as
\begin{eqnarray}
{\cal R}_{MN}= {1\over 2} (T_{MN}^2)_{PQ} \Gamma^{PQ}
+{1\over 4!} (T^4_{MN})_{Q_1\dots Q_4} \Gamma^{Q_1\dots Q_4}~,
\end{eqnarray}
where
\begin{eqnarray}
(T^2_{MN})_{P_1P_2} &=& \tfrac{1}{4} R_{M N, P_1P_2}-\tfrac{1}{12}F_{M[P_1}{}^{Q_1Q_2Q_3}F_{|N|P_2]Q_1Q_2Q_3}~,\notag\\
(T^4_{MN})_{P_1 \ldots P_4} &=& \tfrac{i}{2}D_{[M}F_{N]P_1\ldots P_4}+\tfrac{1}{2}F_{M N Q_1Q_2 [P_1}F_{P_2 P_3 P_4]}{}^{Q_1 Q_2}~ \ .
\end{eqnarray}
The $T^2$ and $T^4$ tensors satisfy various algebraic constraints,
following from the Bianchi identities and field equations:
\begin{eqnarray}
\label{tcon}
(T^2_{MN})_{P_1P_2} &=& (T^2_{P_1P_2})_{MN}~, \cr
(T^2_{M[P_1})_{P_2P_3]} &=& 0~, \cr
(T^2_{MN})_{P}{}^N &=& 0 ~, \cr
(T^4_{[P_1P_2})_{P_3 P_4 P_5 P_6]} &=& 0 \cr
(T^4_{MN})_{P_1P_2P_3}{}^N &=& 0 ~, \cr
(T^4_{M[P_1})_{P_2 P_3 P_4 P_5]} &=& -{1 \over 5!}
\epsilon_{P_1 P_2 P_3 P_4 P_5}{}^{Q_1 Q_2 Q_3 Q_4 Q_5}
(T^4_{M[Q_1})_{Q_2 Q_3 Q_4 Q_5]}~ ,
\end{eqnarray}
and $(T^4{}_{P_1 (M})_{N) P_2 P_3 P_4}$ is totally antisymmetric in $P_1$, $P_2$, $P_3$, $P_4$.
The integrability conditions of the gravitino Killing spinor equations are equivalent to
\begin{eqnarray}
{\cal R}\epsilon^r=0 \ .
\end{eqnarray}
One can obtain constraints on the tensors $T^2$ and $T^4$ by directly evaluating
these constraints on the basis elements $\eta^i$ and using the
constraints and symmetries of $T^2$, $T^4$. However, it is more
straightforward to analyse the constraints by adapting the method
used in \cite{d11preon} to construct a non-existence theorem for
preons in $D=11$ supergravity. In particular, observe that
the constraint ${\cal R}\epsilon^r=0$ implies that
\begin{eqnarray}
{\cal R}_{MN, ab'}=u_{MN,r} \eta^r_{a} \nu_{b'} + u_{MN} \chi_a \nu_{b'}
\end{eqnarray}
where $u$ are complex valued, and ${\eta^r, \chi}$ is a basis for $\Delta_c^+$.
It is also useful to recall the formula
We also have the formula
\begin{eqnarray}
\psi_a \nu_{b'}=-{1\over16} \sum_{k=0}^2 {1\over (2k)!} B(\psi, \Gamma_{A_1A_2\dots A_{2k}}\nu)
(\Gamma^{A_1A_2\dots A_{2k}})_{ab'}~,
\end{eqnarray}
which holds for any positive chirality spinor $\psi$.
Requiring that the holonomy of the supercovariant connection lie in $SL(16,{\mathbb{C}})$ implies that
\begin{equation}
\label{conhol}
u_{MN} B(\chi, \nu) =0
\end{equation}
which eliminates the contribution to  ${\cal R}_{MN, ab'}$ from $ u_{MN} \chi_a \nu_{b'}$.
Combining these expressions, one obtains
Hence we are left with
\begin{eqnarray}
{\cal R}_{MN, ab'}&=&u_{MN,r} \eta^r_{a} \nu_{b'}
\cr
&=& -{1 \over 16} u_{MN,r} \sum_{k=1}^2 {1\over (2k)!} B(\eta^r, \Gamma_{A_1A_2\dots A_{2k}}\nu)
(\Gamma^{A_1A_2\dots A_{2k}})_{ab'}
\end{eqnarray}
which in turn relates $T^2$, $T^4$ to $u_{MN,r}$ via
\begin{eqnarray}
(T^2_{MN})_{A_1 A_2} &=& -{1 \over 16} u_{MN,r} B(\eta^r, \Gamma_{A_1 A_2} \nu)
\cr
(T^4_{MN})_{A_1 A_2  A_3 A_4} &=& -{1 \over 16} u_{MN,r} B(\eta^r, \Gamma_{A_1 A_2 A_3 A_4} \nu) \ .
\end{eqnarray}

To proceed, we relate the components of $T^2$ and $T^4$ to those of $u_{MN,r}$ for
the three different types of canonical normal spinor $\nu$, and then translate the
constraints on $T^2$ and $T^4$ into constraints on $u_{MN,r}$. The analysis
for all possible normals was first constructed in \cite{iibnearmax}, which we also
present here in the remainder of this section.

\subsection{$Spin(7)$-invariant normal}

The normal direction  can be chosen as $\nu = e_5 + e_{12345}$. A suitable basis such that
(\ref{conhol}) is automatically satisfied is
\begin{eqnarray}
\eta^{\bar{\alpha} \bar{\beta}} = e_{\alpha \beta}
~,~~~
\eta^{\bar{\alpha}} &=& e_{\alpha5}~,
\notag\\
\eta^\alpha = {1 \over 6} \epsilon^{\alpha \beta_1 \beta_2 \beta_3}
e_{\beta_1 \beta_2 \beta_3 5} ~,~~~
\eta^+ &=& 1 - e_{1234}~,
\end{eqnarray}
where $\alpha, \beta=1,2,3,4$.
By considering the relation
\begin{equation}
(T^2)_{P_1 P_2} =- {1 \over 16} u_r B(\eta^r, \Gamma_{P_1 P_2} \nu)~,
\end{equation}
where  the form indices $MN$ have been suppressed in $(T^2)$ and in $u_r$,
we find the relations
\begin{eqnarray}
(T^2)_{+-}=(T^2)_{- \mu}= (T^2)_{- \bar{\mu}}= 0~,~~~
(T^2)_{+ \mu} = -{1 \over 8} u_\mu~,~~~
(T^2)_{+ \bar{\mu}} &=& -{1 \over 8} u_{\bar{\mu}}~,
\notag\\
(T^2)_{\mu \nu} = -{1 \over 16} \epsilon_{\mu \nu}{}^{\bar{\beta}_1
\bar{\beta}_2} u_{\bar{\beta}_1 \bar{\beta}_2}~,~~~
(T^2)_{\mu \bar{\nu}} ={1 \over 8} u_+ \delta_{\mu \bar{\nu}}~,~~~
(T^2)_{\bar{\mu} \bar{\nu}} &=& {1 \over 8} u_{\bar{\mu}
\bar{\nu}}~.
\end{eqnarray}
Note that $u_{MN,r}$ are complex valued.
To proceed, observe that
\begin{equation}
u_+ = 2 (T^2)_\alpha{}^\alpha
\end{equation}
and hence, making use of the constraint
$(T^2_{MN})_{P_1P_2} = (T^2_{P_1P_2})_{MN}$,
we find that
\begin{equation}
(T^2_{\alpha \bar{\beta}})_{\mu \bar{\nu}} =
{1 \over 16} (T^2{}_\rho{}^\rho)_\lambda{}^\lambda \delta_{\alpha
\bar{\beta}} \delta_{\mu \bar{\nu}}~.
\end{equation}

Next note that (making use of $(T^2)_{- \mu}=0$)
\begin{equation}
0 = (T^2{}_{N \bar{\beta}})_\mu{}^N = (T^2{}_{\sigma
\bar{\beta}})_\mu{}^\sigma +  (T^2{}_{\bar{\sigma}
\bar{\beta}})_\mu{}^{\bar{\sigma}}~.
\end{equation}
However,

\begin{equation}
 (T^2{}_{\bar{\sigma}
\bar{\beta}})_\mu{}^{\bar{\sigma}} = -{1 \over 16}
\epsilon_\mu{}^{\bar{\beta}_1
\bar{\beta}_2 \bar{\beta}_3} u_{\bar{\beta}_1
\bar{\beta}, \bar{\beta}_2 \bar{\beta}_3}
= -{1 \over 2} \epsilon_\mu{}^{\bar{\beta}_1
\bar{\beta}_2 \bar{\beta}_3} (T^2{}_{\bar{\beta}_1
\bar{\beta}})_{\bar{\beta}_2 \bar{\beta}_3}  =0
\end{equation}
by the Bianchi identity. Hence, it follows that
$ (T^2{}_{\sigma
\bar{\beta}})_\mu{}^\sigma=0$, which implies that
$ (T^2{}_\rho{}^\rho)_\lambda{}^\lambda=0$. Hence
\begin{equation}
(T^2{}_{\alpha \bar{\beta}})_{\mu \bar{\nu}} =0
\end{equation}
so
\begin{equation}
u_{\alpha \bar{\beta},+}=0~.
\end{equation}
Similarly, we also have
\begin{equation}
(T^2{}_{+ \alpha})_{\mu \bar{\nu}} = {1 \over 8} u_{+\alpha,+}
\delta_{\mu \bar{\nu}}
\end{equation}
and hence $u_{+ \alpha,+} = 2 (T^2{}_{+ \alpha})_\lambda{}^\lambda$,
so
\begin{equation}
(T^2{}_{+ \alpha})_{\mu \bar{\nu}} = {1 \over 4}
(T^2{}_{+ \alpha})_\lambda{}^\lambda  \delta_{\mu \bar{\nu}}~.
\end{equation}

Next, note that
\begin{equation}
0 = (T^2{}_{N+}){}_\mu{}^N
= (T^2{}_{\sigma +})_\mu{}^\sigma
+ (T^2{}_{\bar{\sigma} +})_\mu{}^{\bar{\sigma}}~,
\end{equation}
where we have made use of $(T^2)_{+-}=0$.
However, $ (T^2{}_{\bar{\sigma} +})_\mu{}^{\bar{\sigma}}=0$
from the Bianchi identity, hence $ (T^2{}_{\sigma +})_\mu{}^\sigma=0$
also. This implies that $(T^2{}_{+\alpha})_\lambda{}^\lambda=0$, so $(T^2{}_{+\alpha})_{\mu \bar{\nu}}=0$.
Therefore $u_{+\alpha,+}=0$.
Also, $(T^2{}_{+\alpha})_{\mu \bar{\nu}}=0$ implies that $(T^2{}_{+ \bar{\alpha}})_{\mu \bar{\nu}}=0$
(as $T^2$ is real), hence it follows that $u_{+ \bar{\alpha},+}=0$.

The vanishing of $(T^2{}_{\mu \bar{\nu}})_{-\alpha}$, $(T^2{}_{\mu \bar{\nu}})_{- \bar{\alpha}}$,
and $(T^2{}_{\mu \bar{\nu}})_{+-}$ also implies that
$u_{- \alpha,+}=0$, $u_{- \bar{\alpha},+}=0$ and $u_{+-,+}=0$.
Next, consider
\begin{equation}
(T^2{}_{\alpha \beta})_{\mu \bar{\nu}}= {1 \over 8}
u_{\alpha \beta,+} \delta_{\mu \bar{\nu}}~.
\end{equation}
Contracting with $\epsilon^{\alpha \beta \mu}{}_{\bar{\lambda}}$
and using the Bianchi identity we find $u_{\alpha \beta,+}=0$,
so $(T^2{}_{\alpha \beta})_{\mu \bar{\nu}}=0$.
As $T^2$ is real, this implies that
$(T^2{}_{\bar{\alpha} \bar{\beta}})_{\mu \bar{\nu}}=0$, which then fixes
$u_{\bar{\alpha} \bar{\beta},+}=0$.
So all components of $u_+$ vanish.

Next, recall that $(T^2)_{+ \mu}=-{1 \over 8} u_\mu$.
Then the vanishing of $(T^2{}_{+ \mu})_{\alpha \bar{\beta}}$,
$(T^2{}_{+ \mu})_{- \alpha}$, $(T^2{}_{+ \mu})_{- \bar{\alpha}}$ and $(T^2{}_{+ \mu})_{-+}$
implies that
\begin{equation}
u_{\alpha \bar{\beta}, \mu}=0, \qquad u_{- \alpha, \mu}=0, \qquad
u_{- \bar{\alpha}, \mu}=0, \qquad u_{-+, \mu}=0~.
\end{equation}

Next note that
\begin{equation}
(T^2{}_{\alpha \beta})_{+ \mu} = (T^2{}_{+ \mu})_{\alpha \beta}
= -{1 \over 2} \epsilon_{\alpha \beta}{}^{\bar{\rho} \bar{\sigma}}
(T^2{}_{+ \mu})_{\bar{\rho} \bar{\sigma}}~.
\end{equation}
However, we also have $(T^2{}_{+ [\mu})_{\bar{\rho} \bar{\sigma}]}=0$.
Together with $(T^2)_{\mu \bar{\sigma}}=0$ this implies that
$(T^2{}_{+ \mu})_{\bar{\rho} \bar{\sigma}}=0$ and hence
$(T^2{}_{\alpha \beta})_{+ \mu}=0$ also. Hence $u_{\alpha \beta, \mu}=0$.
Furthermore, $(T^2{}_{\bar{\rho} \bar{\sigma}})_{+ \mu}=0$
implies that $u_{\bar{\alpha} \bar{\beta}, \mu}=0$ as well.

Next consider $(T^2)_{+ \bar{\mu}}=-{1 \over 8} u_{\bar{\mu}}$. The
vanishing of $(T^2{}_{+ \bar{\mu}})_{\alpha \bar{\beta}}$,
$(T^2{}_{+ \bar{\mu}})_{- \alpha}$, $(T^2{}_{+ \bar{\mu}})_{- \bar{\alpha}}$,
$(T^2{}_{+ \bar{\mu}})_{-+}$, $(T^2{}_{+ \bar{\mu}})_{\alpha \beta}$
and $(T^2{}_{\bar{\alpha} \bar{\beta}})_{+ \bar{\mu}}$
implies that
\begin{equation}
u_{\alpha \bar{\beta}, \bar{\mu}}=0, \quad
u_{- \alpha, \bar{\mu}}=0, \quad u_{- \bar{\alpha}, \bar{\mu}}=0,
\quad  u_{-+, \bar{\mu}}=0, \quad u_{\alpha \beta, \bar{\mu}}=0, \quad
u_{\bar{\alpha} \bar{\beta}, \bar{\mu}}=0~.
\end{equation}

Next consider the constraint $(T^2)_{\bar{\mu} \bar{\nu}}={1 \over 8}
u_{\bar{\mu} \bar{\nu}}$.
As
\begin{equation}
(T^2{}_{\alpha {\bar{\beta}}})_{\bar{\mu} \bar{\nu}}=
(T^2{}_{\bar{\mu} \bar{\nu}})_{\alpha \bar{\beta}}=0~,
\end{equation}
it follows that $u_{\alpha \bar{\beta}, \bar{\mu} \bar{\nu}}=0$.
Similarly, the vanishing of $(T^2{}_{\bar{\mu} \bar{\nu}})_{- \alpha}$,
$(T^2{}_{\bar{\mu} \bar{\nu}})_{- \bar{\alpha}}$ ,
$(T^2{}_{\bar{\mu} \bar{\nu}})_{+-}$,
$(T^2{}_{\bar{\mu} \bar{\nu}})_{+ \alpha}$
and $(T^2{}_{\bar{\mu} \bar{\nu}})_{+ \bar{\alpha}}$
implies that
\begin{equation}
u_{- \alpha, \bar{\mu} \bar{\nu}}=0, \quad
u_{- \bar{\alpha}, \bar{\mu} \bar{\nu}}=0, \quad
u_{+-, \bar{\mu} \bar{\nu}}=0, \quad
u_{+ \alpha, \bar{\mu} \bar{\nu}}=0, \quad
u_{+ \bar{\alpha}, \bar{\mu} \bar{\nu}}=0~.
\end{equation}

Next consider the Bianchi identity
\begin{equation}
(T^2{}_{\alpha [\beta})_{\bar{\mu} \bar{\nu}]}=0~.
\end{equation}
As $u_+=0$, it follows that
$(T^2{}_{\alpha \bar{\nu}})_{\beta \bar{\nu}}=0$,
and hence $(T^2{}_{\alpha \beta})_{\bar{\mu} \bar{\nu}}=0$.
Therefore $u_{\alpha \beta, \bar{\mu} \bar{\nu}}=0$.
Also
\begin{equation}
(T^2_{\bar{\alpha} \bar{\beta}})_{\bar{\mu} \bar{\nu}}
= -{1 \over 2} \epsilon_{\bar{\mu} \bar{\nu}}{}^{\lambda_1
\lambda_2} (T^2{}_{\bar{\alpha} \bar{\beta}})_{\lambda_1 \lambda_2}
=0~,
\end{equation}
so $u_{\bar{\alpha} \bar{\beta}, \bar{\mu} \bar{\nu}}=0$.
Hence all components of $u_{\bar{\mu} \bar{\nu}}$ vanish.

To summarize, these constraints fix all components of $u_r$ to vanish,
with the exception of $u_{+A,B}$ where $A, B$ are $su(4)$
indices. As
\begin{equation}
\label{spin7con1}
(T^2{}_{+A})_{+B} = -{1 \over 8} u_{+A,B}~,
\end{equation}
it follows that $u_{+A,B}$ is symmetric in $A,B$.

Next consider the 4-forms. It turns out that
all components of $T^4$ are forced to vanish
by the above constraints with the exception of

\begin{eqnarray}
(T^4)_{+\mu \nu \rho} &=& -{1 \over 4} u_{\bar{\alpha}}
\epsilon^{\bar{\alpha}}{}_{\mu \nu \rho}~,~~~
(T^4)_{+ \mu \nu \bar{\rho}} = {1 \over 8} u_\mu
\delta_{\nu \bar{\rho}} - {1 \over 8} u_\nu \delta_{\mu \bar{\rho}}~,
\notag\\
(T^4)_{+ \mu \bar{\nu} \bar{\rho}} &=&
-{1 \over 8} \delta_{\mu \bar{\nu}} u_{\bar{\rho}}
+{1 \over 8} \delta_{\mu \bar{\rho}} u_{\bar{\nu}}
~,~~~
(T^4)_{+ \bar{\mu} \bar{\nu} \bar{\rho}} = -{1 \over 4}
u_\alpha \epsilon^\alpha{}_{\bar{\mu} \bar{\nu} \bar{\rho}}~.
\end{eqnarray}
Using ({\ref{spin7con1}}), this implies that

\begin{eqnarray}
\label{spin7con2}
(T^4)_{+\mu \nu \rho} &=& 2 (T^2)_{+ \bar{\alpha}}
\epsilon^{\bar{\alpha}}{}_{\mu \nu \rho}~,~~~
(T^4)_{+ \mu \nu \bar{\rho}} = (T^2)_{+ \nu} \delta_{\mu \bar{\rho}}
-(T^2)_{+ \mu} \delta_{\nu \bar{\rho}}~,
\notag\\
(T^4)_{+ \bar{\mu} \bar{\nu} \rho}
&=& (T^2)_{+ \bar{\nu}} \delta_{\bar{\mu} \rho}
-(T^2)_{+ \bar{\mu}} \delta_{\bar{\nu} \rho}~,~~~
(T^4)_{+ \bar{\mu} \bar{\nu} \bar{\rho}} = 2 (T^2)_{+ \alpha}
\epsilon^\alpha{}_{\bar{\mu} \bar{\nu} \bar{\rho}}~.
\end{eqnarray}
This implies that $T^4$ is entirely real, so that $F$ is covariantly constant.
Furthermore, $(T^4{}_{+A_1})_{+A_2 A_3 A_4}$
is totally antisymmetric in $A_1, A_2, A_3, A_4$.
Recall that $(T^4{}_{M[P_1})_{P_2 P_3 P_4 P_5]}$
is self-dual in the five anti-symmetrized indices.
Hence $(T^4{}_{+\alpha_1})_{+ \alpha_2 \alpha_3 \alpha_4}$
must vanish. Then ({\ref{spin7con2}}) implies that
$(T^2{}_{+\alpha})_{+ \bar{\beta}}=0$.

Also consider
\begin{equation}
(T^4{}_{+ \alpha}){}_{+ \mu  \nu \bar{\rho}}
= \delta_{\mu \bar{\rho}} (T^2{}_{+ \alpha})_{+ \nu}
- \delta_{\nu \rho} (T^2{}_{+ \alpha})_{+ \mu}~.
\end{equation}
Contracting this identity gives
\begin{equation}
(T^4{}_{+ \alpha})_{+ \mu \lambda}{}^\lambda = -3 (T^2{}_{+ \alpha})_{+ \mu}~.
\end{equation}
However, the self-duality condition implies that
$ (T^4{}_{+ \alpha})_{+ \mu \lambda}{}^\lambda=0$, and hence
$(T^2{}_{+ \alpha})_{+ \beta}=0$ also. Therefore, all components
of $T^2$ and $T^4$ are constrained to vanish.

 \subsection{$SU(4)\ltimes{\mathbb{R}}^8$-invariant normal}

The normal spinor direction is taken to be

\begin{equation}
\nu = (n-\ell +im) e_5 +(n+\ell+im)e_{12345}~,
\end{equation}
and a basis in the space of Killing spinors such that (\ref{conhol}) is satisfied is
\begin{eqnarray}
\eta^{\bar{\alpha} \bar{\beta}} &=& e_{\alpha \beta}~,~~~
\eta^{\bar{\alpha}} =e_{\alpha 5}~,
\notag\\
\eta^\alpha &=& {1 \over 6} \epsilon^{\alpha \beta_1 \beta_2 \beta_3}
e_{\beta_1 \beta_2 \beta_3 5} ~,~~~
\eta^+ = (n-\ell+im)1 - (n+\ell+im)e_{1234}~.
\end{eqnarray}
$T^2$ is constrained by
\begin{eqnarray}
\label{su4t2}
(T^2)_{+-} &=&(T^2)_{- \mu}=(T^2)_{- \bar{\mu}} = 0~,
\notag\\
(T^2)_{+ \mu} &=& -{1 \over 8} (n-\ell +im) u_\mu~,~~~
(T^2)_{+ \bar{\mu}} = -{1 \over 8} (n+\ell+im) u_{\bar{\mu}}~,
\notag\\
(T^2)_{\mu \nu} &=& -{1 \over 16} (n-\ell +im) \epsilon_{\mu \nu}{}^{\bar{\beta}_1
\bar{\beta}_2} u_{\bar{\beta}_1 \bar{\beta}_2}~,~~~
(T^2)_{\mu \bar{\nu}} = {1 \over 8} \big( (n+im)^2-\ell^2 \big)  u_+ \delta_{\mu \bar{\nu}}~,
\notag\\
(T^2)_{\bar{\mu} \bar{\nu}} &=& {1 \over 8} (n+\ell +im) u_{\bar{\mu}
\bar{\nu}}~.
\end{eqnarray}
The analysis proceeds depending on whether or not $ (n+im)^2-\ell^2$ vanishes. There are three cases
but two of them are related by a $Spin(9,1)$ transformation. So there are two
independent cases to consider.

\subsubsection{Generic solutions ($ (n+im)^2-\ell^2 \neq 0$)}

In this case there are no restrictions on the spacetime functions $n,m$ and $\ell$.
It is then straightforward to see,
using the same reasoning as in the $Spin(7) \ltimes{\mathbb{R}}^8$ analysis, that all components of
$u_r$ vanish except for $u_{+A,B}$, where $A=(\alpha, \bar\alpha)$, $B=(\beta, \bar\beta)$, and

\begin{eqnarray}
(T^2{}_{+ \alpha})_{+ \beta} &=& -{1 \over 8} (n-\ell+im) u_{+\alpha,\beta}~,~~~
(T^2{}_{+ \alpha})_{+ \bar{\beta}} = -{1 \over 8} (n+\ell+im) u_{+\alpha,\bar{\beta}}~,
\notag\\
(T^2{}_{+ \bar{\alpha}})_{+ \beta} &=& -{1 \over 8} (n-\ell+im) u_{+\bar{\alpha},\beta}~,~~~
(T^2{}_{+ \bar{\alpha}})_{+ \bar{\beta}} = -{1 \over 8} (n+\ell+im) u_{+\bar{\alpha},\bar{\beta}}~.
\end{eqnarray}
Similarly,  it turns out that
all components of $T^4$ are forced to vanish
by the above constraints with the exception of
\begin{eqnarray}
(T^4)_{+\mu \nu \rho} &=& -{1 \over 4}(n-\ell+im) u_{\bar{\alpha}}
\epsilon^{\bar{\alpha}}{}_{\mu \nu \rho}~,
\notag\\
(T^4)_{+ \mu \nu \bar{\rho}} &=& {1 \over 8} (n-\ell+im) \big( u_\mu
\delta_{\nu \bar{\rho}} - u_\nu \delta_{\mu \bar{\rho}} \big)~,
\notag\\
(T^4)_{+ \mu \bar{\nu} \bar{\rho}} &=&
{1 \over 8} (n+\ell+im) \big( \delta_{\mu \bar{\rho}} u_{\bar{\nu}}
-\delta_{\mu \bar{\nu}} u_{\bar{\rho}} \big)~,
\notag\\
(T^4)_{+ \bar{\mu} \bar{\nu} \bar{\rho}} &=& -{1 \over 4}(n+\ell+im)
u_\alpha \epsilon^\alpha{}_{\bar{\mu} \bar{\nu} \bar{\rho}}~.
\end{eqnarray}

As $(T^4{}_{+A_1})_{+A_2 A_3 A_4}$ is totally antisymmetric in $A_i$,
self-duality implies that $(T^4{}_{+\alpha})_{+\beta \rho \sigma}=0$,
and hence $u_{+\alpha, \bar{\beta}}=0$. Therefore $(T^2{}_{+\alpha})_{+ \bar{\beta}}=0$,
and hence $(T^2{}_{+ \bar{\alpha}})_{+ \beta}=0$ also implies $u_{+\bar{\alpha},\beta}=0$.

Furthermore, we also have
\begin{equation}
(T^4{}_{+\mu})_{+\alpha \beta}{}^\beta = {3 \over 8} (n-\ell+im) u_{+\mu, \alpha}~.
\end{equation}
As the left-hand side of this expression must vanish by self-duality, we find
$u_{+\alpha,\beta}=0$. Hence $(T^2{}_{+\alpha})_{+\beta}=0$, and so
$(T^2{}_{+\bar{\alpha}})_{+\bar{\beta}}=0$ also implies that $u_{+\bar{\alpha},\bar{\beta}}=0$.
Therefore all components of the $u_r$ vanish, so all components of $T^2$ and $T^4$ are constrained
to vanish as well.

\subsubsection{Pure spinor solution ($ (n+im)^2-\ell^2 = 0$)}

There are two pure spinor cases that one can consider depending on whether
 $m=0$, $n=\ell\not=0$ or  $m=0$, $n=-\ell\not=0$.  The normal directions are either $\nu=e_{1234}$ or $\nu=1$, respectively.
 However, these two normals are related by a $Spin(9,1)$ transformation. So it suffices to consider one of the two cases as
 the other will follow by virtue of the $Spin(9,1)$ gauge symmetry of the  Killing spinor equations.
So let us investigate the case
 $m=0$, $n=\ell$. Then ({\ref{su4t2}}) implies that $(T^2)_{+ \alpha}=0$.
Therefore, $(T^2)_{+ \bar{\alpha}}=0$, so $u_{\bar{\alpha}}=0$. Furthermore,
$(T^2)_{\alpha \beta}=0$, so $(T^2)_{\bar{\alpha} \bar{\beta}}=0$ also, and therefore
$u_{\bar{\alpha} \bar{\beta}}=0$. These constraints are sufficient to fix $T^2=0$, however
$u_+$ and $u_\alpha$ are not fixed by constraints involving $T^2$.

It is straightforward to see that the only non-vanishing components of $T^4$ are given by
\begin{eqnarray}
(T^4)_{+ \bar{\alpha} \bar{\beta} \bar{\lambda}} &=& {n \over 2}
\epsilon_{ \bar{\alpha} \bar{\beta} \bar{\lambda}}{}^\rho u_\rho~,~~~
(T^4)_{\bar{\alpha} \bar{\beta} \bar{\lambda} \bar{\sigma}} = -n^2 u_+
\epsilon_{\bar{\alpha} \bar{\beta} \bar{\lambda} \bar{\sigma}}~.
\end{eqnarray}

To proceed, note that the self-duality constraint fixes $(T^4{}_{+ \bar{\sigma}})_{+ \bar{\alpha} \bar{\beta} \bar{\lambda}}=0$,
so $u_{+ \bar{\beta},\alpha}=0$.
Also, $(T^4{}_{+ \sigma})_{+ \bar{\alpha} \bar{\beta} \bar{\lambda}}=-
(T^4{}_{+ \bar{\alpha}})_{+\sigma \bar{\beta} \bar{\lambda}}=0$, so $u_{+ \beta, \alpha}=0$.
Furthermore $(T^4{}_{[\mu \nu})_{\bar{\alpha} \bar{\beta} \bar{\lambda} \bar{\sigma}]}=0$
which implies $(T^4{}_{\mu \nu})_{\bar{\alpha} \bar{\beta} \bar{\lambda} \bar{\sigma}}=0$
and hence $u_{\mu \nu,+}=0$.
Also, $(T^4{}_{[- \nu})_{\bar{\alpha} \bar{\beta} \bar{\lambda} \bar{\sigma}]}=0$
implies $(T^4{}_{- \nu})_{\bar{\alpha} \bar{\beta} \bar{\lambda} \bar{\sigma}}=0$,
so $u_{-\alpha,+}=0$.

Next, consider the following relation implied by self-duality:
\begin{equation}
(T^4{}_{+ [\nu})_{\bar{\alpha} \bar{\beta} \bar{\lambda} \bar{\sigma}]}
= -{1 \over 6} \epsilon_{\bar{\alpha} \bar{\beta} \bar{\lambda} \bar{\sigma}}
\epsilon_\nu{}^{\bar{\lambda}_1 \bar{\lambda}_2 \bar{\lambda}_3}
(T^4{}_{+[-})_{+ \bar{\lambda}_1 \bar{\lambda}_2 \bar{\lambda}_3]}~.
\end{equation}
This implies that
\begin{equation}
n u_{+ \alpha,+} = -{1 \over 2} u_{+-,\alpha}~.
\end{equation}
However, $(T^4{}_{+-})_{+ \bar{\lambda}_1 \bar{\lambda}_2 \bar{\lambda}_3}
= -(T^4{}_{+ \bar{\lambda}_1})_{+- \bar{\lambda}_2 \bar{\lambda}_3} =0$,
which implies that $u_{+-,\alpha}=0$, so $u_{+ \alpha,+}=0$ as well.
Also, $(T^4{}_{[- \rho})_{+ \bar{\alpha} \bar{\beta} \bar{\lambda}]}=0$,
which implies $(T^4{}_{- \rho})_{+ \bar{\alpha} \bar{\beta} \bar{\lambda}}=0$
and so $u_{- \alpha, \beta}=0$.

Also note that $(T^4{}_{- (\bar{\alpha}})_{\bar{\beta}) \bar{\rho} \bar{\sigma}
\bar{\lambda}}= -(T^4{}_{\bar{\rho} (\bar{\alpha}})_{\bar{\beta}) - \bar{\sigma} \bar{\lambda}}=0$,
so
\begin{equation}
u_{- \bar{\alpha},+} \epsilon_{\bar{\beta} \bar{\rho} \bar{\sigma} \bar{\lambda}}
+ u_{- \bar{\beta},+} \epsilon_{\bar{\alpha} \bar{\rho} \bar{\sigma} \bar{\lambda}}=0~.
\end{equation}
Contracting this expression with $\epsilon^{\bar{\beta} \bar{\rho} \bar{\sigma} \bar{\lambda}}$
yields $u_{- \bar{\alpha},+}=0$.

Next consider $(T^4{}_{- (+})_{\bar{\alpha}) \bar{\beta} \bar{\lambda} \bar{\sigma}}
= - (T^4{}_{\bar{\beta} (+})_{\bar{\alpha}) - \bar{\lambda} \bar{\sigma}}=0$.
This implies that
\begin{equation}
n^2 u_{-+,+} \epsilon_{\bar{\alpha} \bar{\beta} \bar{\lambda} \bar{\sigma}}
-{n \over 2} u_{- \bar{\alpha}, \rho} \epsilon_{\bar{\beta} \bar{\lambda} \bar{\sigma}}{}^\rho =0
\end{equation}
and on contracting with $\epsilon^{\bar{\beta} \bar{\lambda} \bar{\sigma}}{}_\mu$, we find
\begin{equation}
\label{su4aux1}
u_{- \bar{\alpha},\mu} = -2n \delta_{\bar{\alpha} \mu} u_{-+,+}~.
\end{equation}
However, self-duality implies that $(T^4{}_{-[+})_{\bar{\alpha} \bar{\beta} \bar{\lambda}
\bar{\sigma}]}=0$, which when combined with ({\ref{su4aux1}}) is sufficient to constrain
$u_{-+,+}=0$ and hence $u_{- \bar{\alpha},\mu}=0$ as well.

Next, note that $(T^4{}_{\mu (\bar{\nu}})_{\bar{\alpha}) \bar{\beta} \bar{\lambda} \bar{\rho}}
= - (T^4{}_{\bar{\beta} (\bar{\nu}})_{\bar{\alpha}) \mu \bar{\lambda} \bar{\rho}}=0$,
hence
\begin{equation}
u_{\mu \bar{\nu},+} \epsilon_{\bar{\alpha} \bar{\beta} \bar{\lambda} \bar{\rho}}
+ u_{\mu \bar{\alpha},+} \epsilon_{\bar{\nu} \bar{\beta} \bar{\lambda} \bar{\rho}} =0~.
\end{equation}
On contracting this identity with $\epsilon^{\bar{\alpha} \bar{\beta} \bar{\lambda} \bar{\rho}}$
we find $u_{\mu \bar{\nu},+}=0$.

The constraint $(T^4{}_{+ (\bar{\mu}})_{\bar{\alpha}) \bar{\beta} \bar{\lambda} \bar{\sigma}}
= - (T^4{}_{\bar{\beta} ({\bar{\mu}}})_{\bar{\alpha}) + \bar{\lambda} \bar{\sigma}}$
implies, on contracting with $\epsilon^{\bar{\alpha} \bar{\beta} \bar{\lambda}
\bar{\sigma}}$, that
\begin{equation}
\label{su4aux2}
6n u_{+ \bar{\mu},+} = - \delta^{\rho \bar{\beta}} u_{\bar{\beta} \bar{\mu}, \rho}
\end{equation}
and furthermore the self-duality constraint $(T^4{}_{\bar{\mu} [+})_{\bar{\alpha} \bar{\beta}
\bar{\lambda} \bar{\sigma}]}=0$ implies, on contracting with
 $\epsilon^{\bar{\alpha} \bar{\beta} \bar{\lambda}
\bar{\sigma}}$, that
\begin{equation}
24 n^2 u_{+ \bar{\mu},+} -12n \delta^{\rho \bar{\beta}} u_{\bar{\beta} \bar{\mu}, \rho}=0~.
\end{equation}
This constraint, together with ({\ref{su4aux2}}) implies that
$u_{+, \bar{\mu},+}=0$ and $\delta^{\rho \bar{\beta}} u_{\bar{\beta} \bar{\mu}, \rho}=0$.
Next note that $(T^4{}_{\bar{\mu} (\bar{\nu}})_{\bar{\alpha}) \bar{\beta} \bar{\rho} \bar{\sigma}}
= - (T^4{}_{\bar{\beta} (\bar{\nu}})_{\bar{\alpha}) \bar{\mu} \bar{\rho} \bar{\sigma}}$.
Contracting this constraint with $\epsilon^{\bar{\alpha} \bar{\beta} \bar{\rho} \bar{\sigma}}$
gives $u_{\bar{\mu} \bar{\nu},+}=0$.

Combining all of these constraints fixes all components of $u_+$ to vanish. To fix the
remaining components of $u_{\alpha}$, note that $(T^4{}_{\bar{\mu} (\bar{\nu}})_{+) \bar{\alpha}
\bar{\beta} \bar{\lambda}} = -(T^4{}_{\bar{\alpha} (\bar{\nu}})_{+) \bar{\mu} \bar{\beta} \bar{\lambda}}$
implies that
\begin{equation}
\epsilon_{\bar{\alpha} \bar{\beta} \bar{\lambda}}{}^\rho u_{\bar{\mu} \bar{\nu}, \rho}
= - \epsilon_{\bar{\mu} \bar{\beta} \bar{\lambda}}{}^\rho u_{\bar{\alpha} \bar{\nu}, \rho}
\end{equation}
and on contracting this expression with $\epsilon^{\bar{\alpha} \bar{\beta} \bar{\lambda}}{}_\sigma$
and using the constraint  $\delta^{\rho \bar{\beta}} u_{\bar{\beta} \bar{\mu}, \rho}=0$
which we have already obtained, we find $u_{\bar{\mu} \bar{\nu}, \sigma}=0$.

Next, note that the constraint $(T^4{}_{\mu (\bar{\nu}})_{+) \bar{\alpha} \bar{\beta} \bar{\lambda}}
= -(T^4{}_{\bar{\alpha} (\bar{\nu}})_{+) \mu \bar{\beta} \bar{\lambda}} =0$
together with $u_+=0$ implies that $(T^4{}_{\mu \bar{\nu}})_{+ \bar{\alpha} \bar{\beta} \bar{\lambda}}=0$,
so $u_{\mu \bar{\nu}, \rho}=0$.
Finally, $(T^4{}_{\mu (\nu})_{+) \bar{\alpha} \bar{\beta} \bar{\lambda}}
= - (T^4{}_{\bar{\alpha} (\nu})_{+) \mu \bar{\beta} \bar{\lambda}} =0$
together with $u_+=0$ imply that $(T^4{}_{\mu \nu})_{+ \bar{\alpha} \bar{\beta} \bar{\lambda}}=0$,
so $u_{\mu \nu, \rho}=0$.

These constraints are then sufficient to fix $u_\alpha=0$, and hence all components of $u_r$
vanish, as do $T^2$ and $T^4$.

\subsection{$G_2$-invariant normal}

The normal spinor can be  chosen as
\begin{equation}
\nu = n (e_5+e_{12345})+im (e_1+e_{234})~.
\end{equation}
By using a gauge transformation of the form $e^{f \Gamma_{+-}}$
for real $f$, we can without loss of generality set $m=\pm n$,
and so we take the normal spinor direction as
\begin{equation}
\nu = e_5+e_{12345} \pm i (e_1+e_{234})~.
\end{equation}
A basis of   spinors compatible with (\ref{conhol}) is
\begin{eqnarray}
\eta^- &=& e_{15}+e_{2345} \mp i (1+e_{1234})~,~~~
\eta^+ = 1-e_{1234}~,
\notag\\
\eta^1 &=& e_{15}-e_{2345}~,~~~
\eta^{1 \bar{p}} = e_{1p}~,~~~\eta^{1p} ={1 \over 2} \epsilon_{pqr} e_{qr}~,
\notag\\
\eta^{\bar{p}} &=&e_{p5}~,~~~
\eta^p = {1 \over 2} \epsilon_{pqr} e_{qr} \wedge e_{15}~,
\end{eqnarray}
where $p,q,r=1,2,3$.
We then find the following constraints on $T^2$:

\begin{eqnarray}
(T^2)_{+-} &=& \pm {i \over 4} u_-~,~~~
(T^2)_{+1} = -{1 \over 8}(u_- - u_1)~,~~~
(T^2)_{+ \bar{1}} = -{1 \over 8} (u_-+u_1)~,
\notag\\
(T^2)_{+p} &=& {1 \over 8} u_p~,~~~
(T^2)_{+ \bar{p}} = -{1 \over 8} u_{\bar{p}}~,~~~
\notag\\
(T^2)_{-1} &=& -{1 \over 8}(-u_- \mp i u_+)~,~~~
(T^2)_{- \bar{1}} = -{1 \over 8}(-u_- \pm i u_+)~,~~~
(T^2)_{-p} = \pm {i \over 8} u_{1p}~,
\notag\\
(T^2)_{- \bar{p}} &=& \mp {i \over 8} u_{1 \bar{p}}~,
\end{eqnarray}

\begin{eqnarray}
\label{g2con1}
(T^2)_{1 \bar{1}} &=& -{1 \over 8}(\pm i u_1 - u_+)~,~~~
(T^2)_{1p} =-{1 \over 8} u_{1p}~,~~~
(T^2)_{1 \bar{p}} =\mp {i \over 8}u_{\bar{p}}~,
\notag\\
(T^2)_{\bar{1} p} &=& \pm {i \over 8} u_p~,~~~
(T^2)_{\bar{1} \bar{p}} = {1 \over 8} u_{1 \bar{p}}~,
\notag\\
(T^2)_{pq} &=& -{1 \over 8} \epsilon_{pq}{}^{\bar{r}}(
u_{1 \bar{r}} \pm i u_{\bar{r}})~,~~~
(T^2)_{p \bar{q}} = -{1 \over 8} \delta_{p \bar{q}}(-u_+ \mp i u_1)~,
\notag\\
(T^2)_{\bar{p} \bar{q}} &=& -{1 \over 8} \epsilon_{\bar{p} \bar{q}}{}^r
(- u_{1r} \mp i u_r)~.
\end{eqnarray}
These constraints imply that
\begin{eqnarray}
\label{g2con2}
u_- &=& \mp 4i (T^2)_{+-}~,~~~
u_1 = -4((T^2)_{+ \bar{1}}-(T^2)_{+1})~,~~~
u_p = 8 (T^2)_{+p}~,
\notag\\
u_{\bar{p}} &=& -8 (T^2)_{+ \bar{p}}~,~~~
u_+ = \pm 4i ((T^2)_{- \bar{1}}-(T^2)_{-1})~,~~~
u_{1p} = -8 (T^2)_{1p}~,
\notag\\
u_{1 \bar{p}} &=& 8 (T^2)_{\bar{1} \bar{p}}~.
\end{eqnarray}
Substituting ({\ref{g2con2}}) back into ({\ref{g2con1}}) gives the
constraints
\begin{eqnarray}
(T^2)_{+1} + (T^2)_{+ \bar{1}} &=& \pm i (T^2)_{+-}~,~~~
(T^2)_{-1} + (T^2)_{- \bar{1}} = \mp i (T^2)_{+-}~,
\notag\\
(T^2)_{-p} &=& \mp i (T^2)_{1p}~,~~~
(T^2)_{- \bar{p}} = \mp i (T^2)_{\bar{1} \bar{p}}~,
\notag\\
(T^2)_{1 \bar{1}} &=& \pm {i \over 2}\big( (T^2)_{+ \bar{1}}
-(T^2)_{+1} + (T^2)_{- \bar{1}} - (T^2)_{-1} \big)~,
\end{eqnarray}
\begin{eqnarray}
(T^2)_{1 \bar{p}} &=& \pm i (T^2)_{+ \bar{p}}~,~~~
(T^2)_{\bar{1} p} = \pm i (T^2)_{+p}~,
\notag\\
(T^2)_{pq} &=& \epsilon_{pq}{}^{\bar{r}} (-(T^2)_{\bar{1} \bar{r}}
\pm i (T^2)_{+ \bar{r}})~,
\notag\\
(T^2)_{p \bar{q}} &=& \pm {i \over 2} \delta_{p \bar{q}}
\big( (T^2)_{- \bar{1}} - (T^2)_{-1} - (T^2)_{+ \bar{1}}
+ (T^2)_{+1} \big)~,
\notag\\
(T^2)_{\bar{p} \bar{q}} &=& \epsilon_{\bar{p} \bar{q}}{}^r
(-(T^2)_{1r} \pm i (T^2)_{+r})~.
\end{eqnarray}
By taking the complex conjugate of these expressions, and using the fact that $T^2_{MN}$ is real, one immediately finds that all components of $T^2_{MN}$ must vanish. This implies, through (\ref{g2con2}), that all components of $u_r$ vanish, and therefore all components of $T^4$ vanish as well.

\section{Solutions with $N=29$ Killing Spinors}

The analysis of solutions preserving 29/32 of the supersymmetry is straightforward. 
First, the results of \cite{homogeneous1} imply that for such solutions, $P=0$.
With this constraint, the algebraic constraint ({\ref{dileq}}) is linear over $\mathbb{C}$,
and so, if ({\ref{dileq}}) admits 29 linearly independent solutions, it must admit
30 linearly independent solutions. Then, by the results of the previous section,
it follows that ({\ref{dileq}}) implies that $G=0$. Finally, as $P=0, G=0$,
the gravitino Killing spinor equation ({\ref{graveq}}) is also linear over $\mathbb{C}$,
so if it admits 29 linearly independent solutions, it must admit 30 linearly independent solutions,
and hence by the results of the previous section, the solution must be locally isometric to a maximally
supersymmetric solution.
	
\section{Conclusions}

In this paper, we have reviewed the work of \cite{iibpreon} and \cite{iibnearmax}
in which is shown that all solutions of type IIB supergravity preserving 29/32, 30/32 and 31/32 of the
supersymmetry are locally isometric to maximally supersymmetric solutions.
However, in order to entirely exclude the existence of such solutions,
one must also show that such solutions cannot arise via quotients of maximally
supersymmetric solutions. We remark that this can occur in some supergravity theories;
for example, it has been shown that all solutions preserving 3/4 of the supersymmetry
in minimal gauged $N=2$, $D=4$ supergravity are locally isometric to the (unique)
maximally supersymmetric solution $AdS_4$ \cite{fourpreon1}. However, is
is possible to explicitly construct a discrete quotient of $AdS_4$ which breaks
the supersymmetry from maximal to 3/4 \cite{fourpreon2}. However, in the
case of IIB supergravity, the analysis of 31/32 supersymmetric solutions presented
here is sufficient to imply that there are no quotients of maximally supersymmetric
solutions preserving exactly 31/32 of the supersymmetry. This is because
the constraints $P=0$, $G=0$ imply that the Killing spinor equations are linear over
$\mathbb{C}$, and so the space of Killing spinors is even-dimensional.
In fact, it has been shown in \cite{iibnearmax} that there
are also no quotients of maximally supersymmetric solutions
which preserve exactly 30/32 or 29/32 of the supersymmetry. 
In the case of $D=11$ supergravity, it has been shown  
that all 31/32-supersymmetric solutions are locally maximally supersymmetric \cite{d11preon},
and that there are no quotients of maximally supersymmetric solutions which
preserve 31/32 of the supersymmetry \cite{figquot}.

Having established these results, it is natural to attempt to extend the analysis
presented here to include solutions preserving lower proportions of supersymmetry.
In the case of IIB supergravity, it is known that there exists a solution
preserving 28/32 of the supersymmetry. The solution is a plane wave geometry found in
\cite{xxviiisol}. This solution has as expected, $P=0$, with $F \neq 0$, and 
$G \neq 0$. It should be noted that the integrability conditions of IIB supergravity
are significantly more complicated when one has non-vanishing $G$, so we expect
the analysis to be considerably more involved. 
Similar calculations should also be possible in $D=11$ supergravity. 
Analogous homogeneity results have been constructed in $D=11$ supergravity,
it has been proven that all solutions preserving more than 24/32 of the supersymmetry
are homogeneous \cite{homogeneous2}. However, the structure of the Killing spinor
equation of $D=11$ supergravity differs from the Killing spinor equations of IIB supergravity.
There is no purely algebraic Killing spinor equation in $D=11$ supergravity, which could
be readily simplified by making use of homogeneity. One must instead work  directly with the
integrability conditions of the gravitino equation, which have a rather complicated structure.
Nevertheless, it is reasonable to expect that the homogeneity of solutions preserving more than
24/32 of supersymmetry in IIB and $D=11$ supergravity should play an important role in
constructing classifications of these solutions, it would be interesting to obtain these classifications.

\section*{Acknowledgements}

J.G. would like to thank the organizers of the ``Special Metrics and
Supersymmetry'' conference at Universidad del Pa\'is Vasco, May 2008.

\setcounter{section}{0}

\end{document}